# Cultural association based on machine learning for team formation

Hrishikesh Kulkarni[1], Bradly Alicea[2,3]

[1] Computer Engineering Department, PVG's COET, SPPU, Pune, INDIA
hrishikeshparag@gmail.com; [2] Orthogonal Research and Education, Champaign, IL USA; [3] Open Worm Foundation, Boston, MA USA bradly.alicea@outlook.com

**ABSTRACT**

Culture is core to human civilization, and is essential for human intellectual achievements in social context. Culture also influences how humans work together, perform particular task and overall lifestyle and dealing with other groups of civilization. Thus, culture is concerned with establishing shared ideas, particularly those playing a key role in success. Does it impact on how two individuals can work together in achieving certain goals? In this paper, we establish a means to derive cultural association and map it to culturally mediated success. Human interactions with the environment are typically in the form of expressions. Association between culture and behavior produce similar beliefs which lead to common principles and actions, while cultural similarity as a set of common expressions and responses. To measure cultural association among different candidates, we propose the use of a Graphical Association Method (GAM). The behaviors of candidates are captured through series of expressions and represented in the graphical form. The association among corresponding node and core nodes is used for the same. Our approach provides a number of interesting results and promising avenues for future applications.

## INTRODUCTION

What is the reason that a some people can work together exceedingly well while others perform poorly? Every individual is unique and exhibit certain cultural traits. One potential reason is that some individuals share cultural similarities with other individuals. Culture refers to typical behavioral patterns, beliefs and thought process which also includes reasoning. Building a cultural context involves establishing shared ideas. Cultural context is transmitted through education [1], which helps to define a social group defined by these shared ideas. Identifying cultural similarity among candidates can be thought of finding similarity among individuals' established ideas. It is a multifaceted aspect and it always remains challenge to determine cultural similarity. Further impact of cultural similarity can help in determining two or more candidates can perform together.

In this paper, we define culture as a set of typical behaviors, beliefs, and thought process (e.g. reasoning) [2]. Yet cultural similarity is more systemic problem. It considers different aspects about behaviors. Even cortical activation differ with different cultures while doing certain activities like arithmatic processing [3]. Non-parametric path analysis can be used while associating two or more behaviors [4]. Modelling culture has always been area of interest for researchers, traditionally involving researchers from computer engineering, mathematics, and social psychology. Researchers have previously shown a strong relationship between culture, attitude, and expressions [5]. Culture thus closely associated with social behavior. Thus it is reflected in an individual's interactions, decision making and social interactions [6]. Exposure impacts on behavior. This holistic relationship impacts overall interactions [7] [8]. Knowledge of the cultural association and behavioral patterns among candidates can help in forming groups, promoting products and most importantly delivering social messages to groups and individuals. Hence human behavior and culture can help in formation of teams, groups for a mission [9]. These behavioral patterns may either be restricted to a single culture, or cross cultural boundaries [10].

In all this process determining cultural association among candidates and groups can be key to overall success. In this paper, we propose a measure of cultural association through expression captured using multiple-choice questions (MCQs), hybrid distance measurement, and a dynamic variable clustering model. The expressions here are captured by monitoring activities, responses and responses to questions in the context of situations.

This paper is organized in five sections. The Related Work section provides an overview of work carried out in culture modelling, Cultural Computing and Social Psychology. The Proposed Model section provides the details of a machine learning and graph based association model for mapping and clustering based on expressions. The Data Preparation section includes a proposed mathematical model, while the Results and Conclusion sections demonstrate and elaborate upon the results obtained using this model.

**RELATED WORK**
The statistical modelling of culture has gained relevance in recent years. Previous approaches for modelling culture include SVM and random forest algorithms [11]. Typically, culture is captured through unstructured information using methods specialized for analyzing textual and linguistic data.

Expression mining is one such approach that is loosely related to NLP and text processing. Expressions is overall systemic concept and it comes in the form of text, voice, response, interactions and behaviors. But text is one of the core parts of these expressions. Text in news and blogs also convey emotions of the writer. It involves clustering and identifying the most relevant text. The bigrams and occurrence of certain text together can give pointers cultural significance. Though text similarity is not directly related to cultural similarity, it can be coupled with context of occurrence to provide clues to cultural structure. To identify text similarity longest common subsection can be identified. This can even be very useful in plagiarism detection [12].

Cosine similarity is one of the most popular techniques used in identifying textual similarity [13]. When it comes to cultures and behaviors using textual expressions semantic similarity is of prime importance while associating them. The sentiment is also equally important. Hence, sentiment computing can be used while clustering candidates with reference to cultural traits. It involves behavioral analysis and emotional computing [14]. Semantic similarity is also used for sentiment analysis and that can help in deriving overall crux for cultural association. The expression can be represented in terms of textual artifacts typically paragraphs or documents. To associate two or more documents normalized similarity based semantic approach can be used [15]. Semantic vector of words [16, 17] can be calculated by finding out synonyms for overlapping words. Cultural similarity and emotions can even be very relevant while product selection and embedding emotions into products for certain segment of customers [18].

Cultural association needs to consider different parts of expressions. In all these cases behaviour and emotions are important. The emotional reflections can be very well used for wellness [19]. Researchers also worked on deriving emotions based on expressions coming in the form of text. Emotions are derived using single prominent keyword or association of multiple keywords [20, 21, 22].

**PROPOSED MODEL**
Our methodology is based on context determination. It uses extraction of relationships among words with reference to cultural significance. This includes core theme identification. This is derived from core words identification and extracting relationships among words. The distance based association is represented in matrix form which can easily be converted in to graphical representation.

A model needs data. Hence, data acquisition and pre-processing is required before development of model. Since suitable third-party data is not available we focused on capturing relevant data. For data preparation, key questions are identified those can depict cultural traits. These questions are prepared with multiple revision and under supervision of psychologist and computational social science researchers. The answers to these questions are collected from social network interactions and other communications from each candidate.

Self-report surveys are also utilized, based on psychological and anthropological expertise. Twenty-two (22) questions are created with the inputs from the experts, common issues in the literature, and our specific objectives. Responses to questions are captured in the form of images in selected cases and MCQ options or descriptive text in rest of the cases. While preparing data, the following issues were taken into consideration:

- Data collected from various candidates with different social, cultural and economic backgrounds.

- Data collected from candidates from different geographical locations.

- Manual verification of data carried out before entering data into data set. This verification is done to verify authenticity and consistency of data.

- In first phase to check validity of questions and expression-based approach, answers from eight (8) candidates are selected and different experiments are carried out on this initial set.

- In next phase, responses from 100 candidates are captured and next level validation is performed.

To analyze the survey data (MCQs), *k*-means method is used to check manual closeness with Euclidean distance measure. Data preprocessing done for MCQ and Textual inputs. The number of clusters are decided based on the data distributions (elbow and silhouette methods are used for this purpose). This method ensures that the data collected can be clustered based on similarities.

For textual data filtering, Nonlinear Programming (NLP)- based preprocessing was done. A manual association based on MCQ relevance is conducted in the Indian cultural context. These contextual features are then used to generate graphical association method (GAM) models. A Principle Component Analysis (PCA) is then used to get association and their validation between selected features. Finally, an association graphs matrix is generated from MCQ data.

Thus, the proposed algorithm focuses on text data processing to derive cultural association. It consists of four major parts. The algorithm is given below:

- Derive Context: To derive the intent of candidate the answer-text is preprocessed and natural language processing and topic determination is used to determine the topic. These topics are associated to derive the context. The context along and language processing is key to derive intent.

- Association to derive intent: In the next stage similarity among text artifacts and overall intent is determined using hybrid techniques. Two basic methods used for initial association are:

  o Term Distance

  o LSTM dictionary-based word

- Similarity Score: The similarity score is determined using closeness factor. The similarity score and association is used to generate Graphs. In these Graphs weights are generated for text inputs
  - LSTM based word association is used to create weighted graph
  - WordNet association is considered for the same
- Cultural association: The overall distance calculation is done based on Google news vectors. Each candidate's cultural parameters are represented as a graph. Every graph has a core theme. The association among graphs is determined node-by-node using the GAM, where the corresponding node among two graphs are determined and with reference to core theme the association among two graphs is determined.

Thus, the overall Analysis is performed using GAMs and data analysis. Association among data can be divided into three parts: MCQ data graphs, textual data graph, and textual similarities. Association of MCQ data in graph with Indian context creates meaningful information to determine the categories of person ranging from rule base, fun loving and many more. These categories are kept limited and are derived from the core word in the graph.

Textual data graph represents association of keywords found by long short-term memory (LSTM) and genesis. It also uses concepts which represent candidate features from the text data. This will ensure the association of concepts for each candidate is mapped using graph similarities with simple Euclidian distance measurement. The overall association derives the cultural context – thus the GAM model helps us to derive cultural traits of individuals.

Text data key vector extraction and similarity shows the association between two textual data. The outcome of cultural association in terms of theme and context (e.g. theme of Indian classical music) is determined by resolving graphs and building the overall context. To achieve these following steps are carried out:

- Cultural similarities are determined using mapping of MCQ data graphs. Thus, a chronological cultural context is set.
- Cultural similarities using textual graph is mapped. These similarities are quantified in numbers.
- Cultural similarities among candidates are determined using text vector similarities

### DATA PREPARATION

We utilize a number of intelligent methods for associating cultural attributes. All machine learning techniques are data hungry and it becomes even more critical when we are dealing with cultural and emotional aspects of humans. Direct questions have their own limitations and can often lead to biased responses. Hence, indirect questionnaires are used to capture cultural aspects of human being: these questions do not attempt to elicit confidential, or sensitive information. These questions are finalized after three revisions under guidance from psychology experts and analysis with reference to problem at hand. While creating questions following major aspects are considered:

- It should be interesting: The candidate should love to answer these questions.
- Easy to answer: Candidate will not have to take pain to answer it.

- Relevant but not direct: The information should be derived from indirect clues since direct answers are mostly misleading.

- Make the candidate think: It should get candidates heart out.

Twenty-two (22) questions are considered which do not contain any personal information. Data is prepared that can help in decoding context based on behavior of candidates. Attribute considered are depicted in Table 1.

Table 1. Attributes to build context.

| Location | Tradition | Religion | Traveling attributes |
|---|---|---|---|
| Behavior attributes | Work Information | Social attributes | Week days and weekend routine |
| Hobbies | Events | | |

Data collected from candidates is basically represented in the form of question and answers. These candidates are purposefully selected from different backgrounds. The data clearly shows that the variations are based on the following factors

a. Locational

b. Personal Preferences

c. Familial information

d. Social and Behavioral

This variation can lead to people with certain personal attributes selecting different options in the MCQs. One example is a person who lives near sea coast but may want to travel to a hilly area. This can be contrasted with people who consistently select travel to regions that are like the one in which they live. This will create two categories for personal attributes: flexible and rigid. Textual and non-textual data can have different degrees of association for different candidates.

The data are filtering and then converted into both numeric and non-numeric data. Filtering is done on text data as stop word removal, stemming and other methods are applied for cleaning and filtering the data. The previously defined *k*-means clustering method is used to validate the thought process of feature creation. Initial data are captured and a *k*-means analysis is conducted. Our analysis then proceeds as follows:

1. Selection of number of clusters is done to express the categories

2. Optimal number of clusters defines number of groups that can be created

3. Optimal number of clusters are selected using elbow and silhouette method

Clusters are formed using numeric MCQ data based on options selected. Let us assume that two people having nearly same traveling and personal attributes are represented by the same cluster. In this case, the distance between them should be minimal (e.g. asymptotic to zero). The following measure represents the sum of intra-cluster distances between points in a given cluster $C_k$ containing $n_k$ points. Adding the normalized intra-cluster sums of squares gives a measure of the compactness of our clustering:

$$W_k = \sum_{k=1}^{K} \frac{1}{2n_k} D_k \qquad (1)$$

This variance quantity $W_k$ is the basis of a naive procedure to determine the optimal number of clusters: the elbow method. To demonstrate this, let us use one example: travel features created manually based on association of combination of column attributes like traveling alone and going at beach will reduce feature weight as show in Table 2.

Table 2. Questions to graph mapping.

| Feature | Travel 1 | Travel 2 | Travel 3 | Travel 4 |
|---|---|---|---|---|
| Questions | 2-1 | 2-3 | 3-1 | 4-5-7 |
| Formula weight (*W*) | *W*= (Q1response-Q2response)/group_ratio | | | *W* = can travel with known or unknown |
| Graph | Combined graph (cyclic or acyclic) | | | Link to combined graph with weight |

In case of textual data, NLP-based similarity is calculated then filtering is done on textual data. Cosine similarity is used and can be defined mathematically as

$$\vec{a} \cdot \vec{b} = \|\vec{a}\| \|\vec{b}\| \cos\theta \qquad (2)$$

$$\cos\theta = \frac{\vec{a} \cdot \vec{b}}{\|\vec{a}\| \|\vec{b}\|} \qquad (3)$$

The context vector is generated using Keras, Genesis and spacy. Here WordNet and Wikipedia based data are used to associate user data to vector. Personalized dictionary is created for designing graph based on the data captured. A context graph is then created and weights are assigned, in addition to generation of a GAM demonstrating a star topology. Each text key vector is connected to other context vectors. Thus, context vector is represented as a matrix given by eq. 4.

$$\begin{bmatrix} W_1 \\ W_2 \\ \\ W_n \end{bmatrix} \begin{bmatrix} f_1 & f_2 & \dots f_n \end{bmatrix} \qquad (4)$$

**RESULTS**

Various methods are used to find potetntial associations. Feature creation and association from MCQ data (shown in matrix) is used for further processing. Manual attribute association and combination of attributes creates a feature. Cultural similarity always talks in group similarity. These similarities can be plotted as graph nodes. Various text similarity methods are applied like cosine similarity, and distance measure. The results show that two candidates may be associated contextually but because as there is no direct text association it is visible we may fail to determine cultural associating between them. Personalized dictionary-based context vectors used for graph generation (shown in matrix). Thus, contextual associating through expression helps to determine cultural association in much conclusive way. A sample context vector is depicted in Table 3.

Table 3. Example context vector.

Vector example ('book', 'series', 'rap', 'people', 'harry', 'snape', 'khatam', 'gulab', 'jamun', 'indian') (0.08781022329052053, 0.08314971582897987, 0.07726249198125512, 0.07468618760293577, 0.07142857142857142, 0.07142857142857142, 0.07142857142857142, 0.07142857142857142, 0.07142857142857142, 0.06976656954940326)

From the similarity graph based on traveling preferences, we can learn several things. First, travel features can be discovered and associated (Table 1).

1. Personal features are created and associated as shown in Table 2.

2. Text key vectors are generated based on the candidate's data, text feature 1 to 9 are associated to context text key vector.

3. Text2vector is generated for entire data and similarity is calculated with other candidate data, this is direct mapping.

The comparison among clusters is depicted in Figures 1 and 2. The figures depict the cluster to which different candidates belong with reference to clustering methods used.

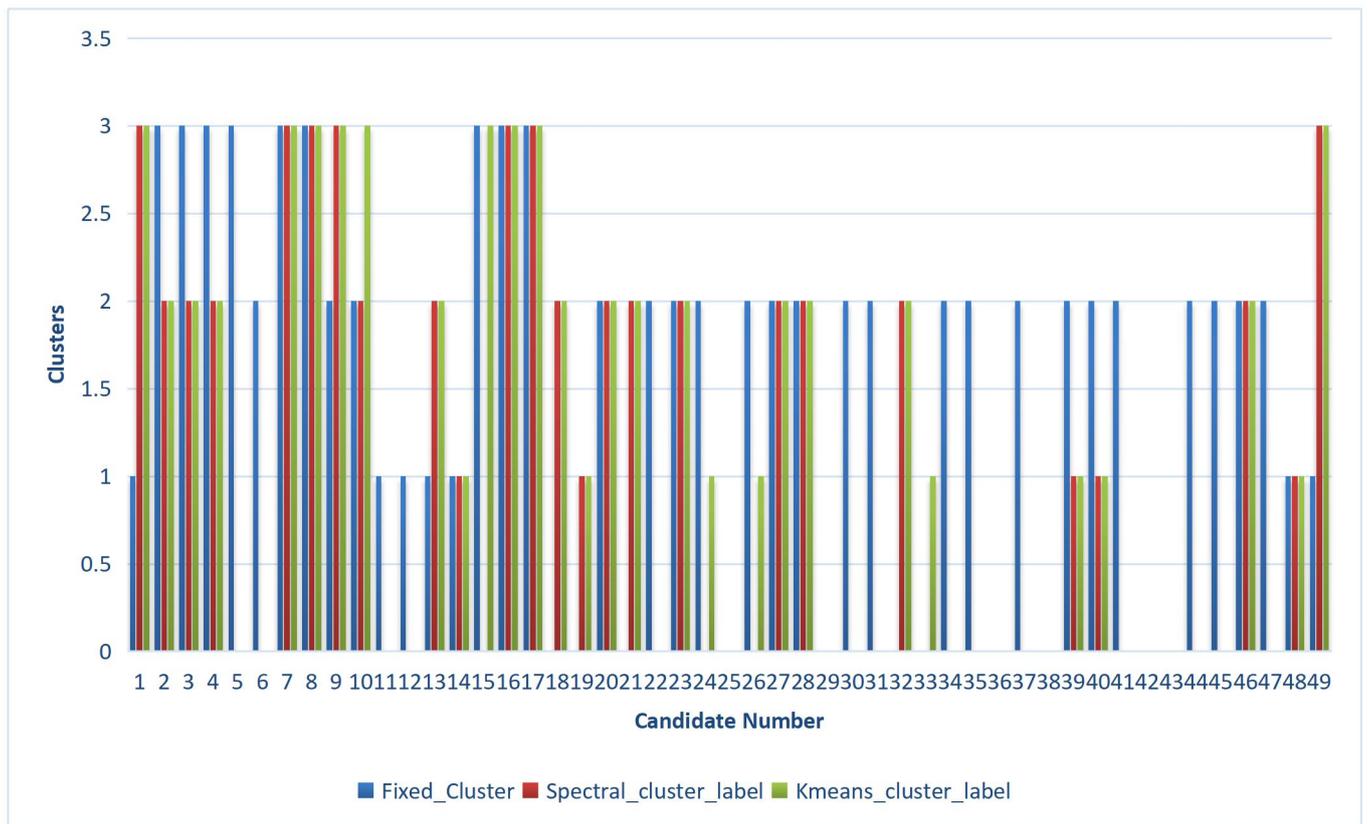

Fig 1. Comparing cluster membership ($k$=3) across candidate pairs for three different clustering methods on the survey data.

These comparisons are suggestive that – in spite of selection of different methods the candidates of similar cultural traits share similar cultural labels. This similarity can be very useful while marketing products, recommending news and forming groups for certain objective. There can be myriad applications of knowing the cultural association beforehand.

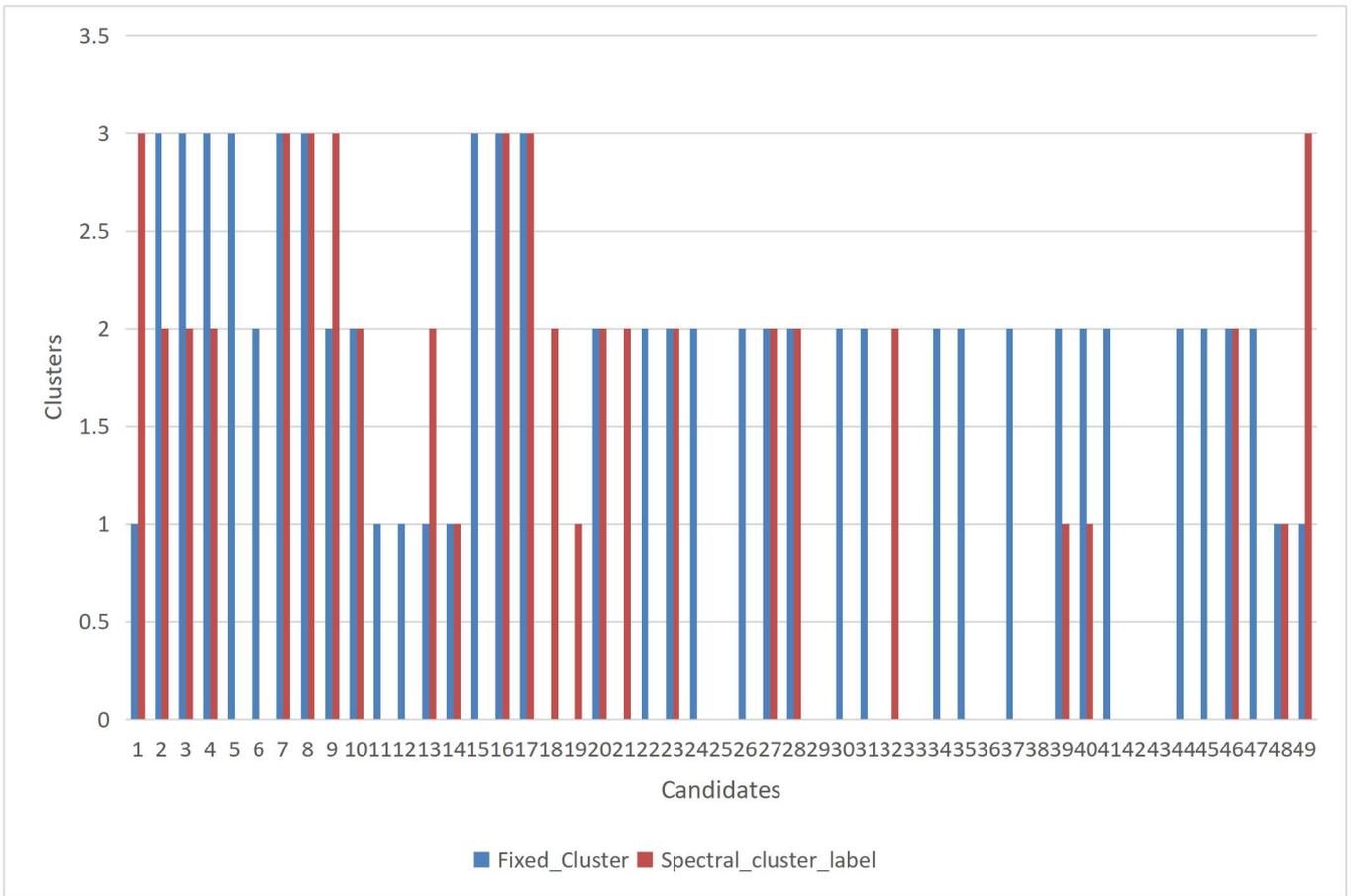

Fig 2. Comparing cluster membership (*k*=3) across candidate pairs for the fixed clustering versus spectral clustering methods.

## CONCLUSION

The cultural associations among candidates shown here can help foster better team formation and partner selection. Clustering candidates based on cultural affinity and behavioral analysis has always been challenge to computer science and cognitive science researchers alike. This paper proposes a graph-based clustering approach to classify individuals based on their cultural attributes. The results are analyzed with reference to data collected for 100 candidates, but the method may be scale to the analysis of thousands of candidates. The results are verified with reference to labeled data. The observed accuracy rate is around 87%, and can be verified with an expert's logical answers. This algorithm can further be improved with multi-level graph clustering and filtering while association among candidates is established. The promising result presented here coupled with the myriad uses of the cultural association approach will aid in developing more sophisticated technologies related to group cognition and culturally-influenced behavior.

## REFERENCES

[1] de Bono, E. (1990). Lateral Thinking, Penguin Books.

[2] Kulkarni, H. & Alicea, B. (2018). Cultural Affinity through Associative Machine Learning and Behavioral Computation. *5th International Psychological Congress (NAPS)*.

[3] Yu, J., Pan, Y., Ang, K.K., Guan, C., Leamy, D.J. (2012). Prefrontal Cortical Activation during Arithmetic Processing Differentiated by Cultures: A Preliminary fNIRS Study. 34th Annual International Conference of the IEEE Engineering in Medicine and Biology Society (EMBC), San Diego, USA.

[4]     Zhang, J. & Bareinboim, E. (2018). Non-Parametric Path Analysis in Structural Causal Models" UAI-18. In *Proceedings of the 34th Conference on Uncertainty in Artificial Intelligence*, 2018. Purdue CausalAI Lab, Technical Report (R-34).

[5]     Mooij, Marieke & Hofstede, Geert. (2011). Cross-Cultural Consumer Behavior: A Review of Research Findings. *Journal of International Consumer Marketing,* 23, 181-192.

[6]     Henrich, J. (2015). Culture and Social Behavior. *Current Opinion in Behavioral Sciences*, 3, 84–89

[7]     Markus, H., & Kitayama, S. (1991). Culture and the self: Implications for cognition, emotion, and motivation. *Psychological Review*, 98, 224-253.

[8]     Zajonc, R.B., Markus, H., & Wilson, W. R. (1974). Exposure effects and associative learning. *Journal of Experimental Social Psychology*, 10, 248-263.

[9]     Kulkarni, H. & Marathe, M. (2019). Context Vector Convergence (CVC) of Computational Behaviour and Cultural Traits for Team Selection. *International Journal of Information and Decision Sciences*. Forthcoming.

[10]    Douglas, S.P. & Craig, C.S. (1997). The changing dynamic of consumer behavior: implications for cross-cultural research. *International Journal of Research in Marketing*, 14, 379-395.

[11]    Leo, B. (2001). Statistical Modeling: The Two Cultures. *Statistical Science*, 16. doi:10.1214/ss/1009213726.

[12]    Alberto, B. Paolo, R. Eneko, A. & Gorka, L. (2010). Plagiarism Detection across Distant Language Pairs. *Proceedings of the International Conference on Computational Linguistics*, 37–45.

[13]    Qian, G., Sural, S., Gu, Y., & Pramanik, S. (2004). Similarity between Euclidean and cosine angle distance for nearest neighbor queries. *Proceedings of ACM Symposium on Applied Computing*, 1232-1237.

[14]    Jiang, D., Luo, X., Xuan, J., & Xu, Z. (2017). Sentiment Computing for the News Event Based on the Social Media Big Data, *IEEE Access*, 5, 2373-2382.

[15]    Sharma, S., Lather, J.S., & Dave, M. (2014). Normalized similarity based semantic approach for discovery of web services. *IEEE International Advance Computing Conference (IACC)*, doi:10.1109/IAdCC.2014.6779369.

[16]    Saad, S.M. & Kamarudin, S.S. (2013). Comparative analysis of similarity measures for sentence level semantic measurement of text. *IEEE International Conference on Control System, Computing, and Engineering*, doi:10.1109/ICCSCE.2013.6719938.

[17]    Mitchell, J., & Lapata, M. (2008). Vector-based Models of Semantic Composition. *Proceedings of the Association for Computational Linguistics*, 236-244.

[18]    Kulkarni, H., Joshi, P., Chande, P.K. (2019). Computational Psychology to Embed Emotions into Advertisements to Develop Emotional Bonding. *Indian Journal of Psychological Science*. Forthcoming.

[19] De Choudhury, M., Gamon, M., Hoff, A., & Roseway, A. (2013). Moon Phrases: A Social Media Facilitated Tool for Emotional Reflection and Wellness. *Proceedings of the 7th International Conference on Pervasive Computing Technologies for Healthcare*, 13679632.

[20] Liu, B. (2010). Sentiment Analysis and Subjectivity. In "Handbook of Natural Language Processing", N. Indurkhya & F.J. Damerau eds. CRC Press, Boca Raton, FL.

[21] Quan, C. & Ren, F. (2011). Selecting clause emotion for sentence emotion recognition. *7th International Conference on Natural Language Processing and Knowledge Engineering*, Tokushima JAPAN.

[22] Aisopos, F., Papadakis, G., Tserpes, K., & Varvarigou, T. (2012). Textual and contextual patterns for sentiment analysis over microblogs. *Proceedings of the 21st International Conference on World Wide Web*, 453-454.